# Exploring Hybrid Work Realities: A Case Study with Software Professionals From Underrepresented Groups

Ronnie de Souza Santos
University of Calgary
Calgary, AB, Canada
ronnie.souzasantos@ucalgary.ca

Cleyton Magalhaes
University Federal Rural de Pernambuco
Recife, PE, Brazil
cleyton.vanut@ufrpe.br

Robson Santos
UNINASSAU
Triunfo, PE, Brazil
robson.rtss@gmail.com

Jorge Correia-Neto
University Federal Rural de Pernambuco
Recife, PE, Brazil
jorgecorreianeto@gmail.com

## ABSTRACT

*Context*. In the post-pandemic era, software professionals resist returning to office routines, favoring the flexibility gained from remote work. Hybrid work structures, then, become popular within software companies, allowing them to choose not to work in the office every day, preserving flexibility, and creating several benefits, including an increase in the support for underrepresented groups in software development. *Goal*. We investigated how software professionals from underrepresented groups are experiencing post-pandemic hybrid work. In particular, we analyzed the experiences of neurodivergents, LGBTQIA+ individuals, and people with disabilities working in the software industry. *Method*. We conducted a case study focusing on the underrepresented groups within a well-established South American software company. *Results*. Hybrid work is preferred by software professionals from underrepresented groups in the post-pandemic era. Advantages include improved focus at home, personalized work setups, and accommodation for health treatments. Concerns arise about isolation and inadequate infrastructure support, highlighting the need for proactive organizational strategies. *Conclusions*. Hybrid work emerges as a promising strategy for fostering diversity and inclusion in software engineering, addressing past limitations of the traditional office environment.

## KEYWORDS
hybrid work, post-pandemic, LGBTQIA+, PWD, Neurodivergent

## 1 INTRODUCTION

The COVID-19 pandemic became one of the most devastating events in the history of humankind. As the virus spread across the globe, it forced society to find alternative ways to keep functioning [41]. In particular, remote work became one of the only alternatives for organizations. In this process, many professionals had to deal with working challenges when adjusting their homes to unconventional daily routines, e.g., working in improvised home offices, sharing the space with family members, or juggling work with childcare [3, 6, 8, 28].

After months of restrictions, when the pandemic was under control, many companies started asking employees to return to their offices; however, professionals adapted to the challenges experienced at the beginning of the pandemic, and the idea of returning to an office routine was no longer attractive [5]. In the software industry, the plans to return to the office triggered different feelings among software engineers. Some of them were keen; others were trepidatious, but most of them were convinced by personal experience that their work could be done remotely [10]. Therefore, they expect to continue working from home some or all the time [29, 38]. This is why several software companies decided to adopt hybrid work structures [7, 40].

Hybrid work is a spectrum of flexible work arrangements [39], in which employees usually belong to the same organization and mostly live in the same region where the office is (i.e., no outsourcing); they simply choose not to work in the office every day [10, 13, 39]. For many professionals, this is a chance to keep the flexibility gained from remote work, such as dealing with personal responsibilities (e.g., parenting) and avoiding commuting [17, 34]. However, for those belonging to underrepresented groups, embracing hybrid work entails breaking down the various barriers they previously faced when working exclusively on-site [11, 12].

In this study, we investigated how software professionals from underrepresented groups are experiencing post-pandemic hybrid work by exploring the case of a large South American software company that adopted a voluntary hybrid-work model, i.e., professionals can decide if and when they are going to the office. Our motivation lies in the need to understand how inclusive hybrid work in software engineering can be for certain groups. Therefore, we explored the experiences of three groups of professionals who belong to underrepresented groups in the software industry:

neurodivergents[1], people with disabilities (PWD)[2], and LGBTQIA+ individuals[3]. To address this topic, we focused on answering the following research question: **RQ.** *How are software professionals from underrepresented groups dealing with hybrid work?*

From this introduction, our study is organized as follows. In Section 2, we present studies about how individuals from underrepresented groups faced remote work. In Section 3, we describe how we conducted the case study, while Section 4 presents our findings. In Section 5, we discuss the implications and limitations of our study. Finally, Section 6 summarizes the contributions of this study.

## 2 BACKGROUND

Remote work refers to a work arrangement where professionals perform their tasks and responsibilities from a location outside the traditional office, typically from home, but also from other remote settings [47]. Although remote work has existed for decades, the COVID-19 pandemic catalyzed a significant increase in its adoption, driven by companies' needs to maintain business continuity while also prioritizing health and safety concerns [44]. Nowadays, in the post-pandemic era, remote work can be seen as one of the dimensions within the spectrum of hybrid work [39], i.e., professionals have the ability to work from the office but decide to work mostly from home, while their co-workers might be in the office.

The flexibility of working outside the office is appealing to a wide range of professionals as the adaptability of this environment has proven to accommodate the different requirements of several individuals, from parents who need to balance caregiving responsibilities to individuals with disabilities seeking a more accessible workspace and those dealing with personal unique needs [24]. In particular, the pandemic underscored the potential of remote work to foster an inclusive and diverse professional environment [16]. Remote work can empower individuals belonging to different underrepresented groups, such as those with disabilities, neurodivergent traits, and LGBTQIA+ community members, by improving inclusive environments through personalized workspaces that cater to individual needs, offering flexibility in scheduling arrangements and reducing biases related to physical appearance [45].

Working from home can be particularly advantageous for people with disabilities by allowing them to establish personalized and accessible work setups at home. Additionally, this work arrangement can support these individuals in mitigating various commuting challenges, including physical barriers, inaccessible transportation, or unreliable public spaces [22, 45]. For neurodivergent individuals, remote work provides personalized workspaces, flexible schedules, and the ability to accommodate diverse communication channels, which could minimize sensory challenges and social anxiety. Yet, without appropriate support, they may face significant cognitive labor in configuring an accessible physical and virtual setup to conduct their tasks [43]. Looking at the LGBTQIA+ community, working from home offers them a private and safe space to express their identity, reducing exposure to potential discrimination in public spaces. This enables these professionals to concentrate on their tasks without constantly being concerned about how they are perceived by their co-workers, particularly due to the increased access to more inclusive virtual work environments [27].

It is worth noticing that despite several reported benefits, the literature also discusses challenges faced by individuals from underrepresented groups in remote working environments. These challenges often stem from potential disparities in resource access and a lack of workplace support, which might affect the development and effectiveness of diversity and inclusion policies in various organizations [1, 4]. In the specific context of the software development industry, research on the impact of remote work on software professionals aligns with general literature findings. Despite challenges, remote work has been predominantly positive for professionals from underrepresented groups, creating opportunities, increasing flexibility, and contributing to a more diverse and inclusive software engineering [9, 11, 12, 32, 34].

Now, software companies are embracing hybrid work, which, according to recent studies in software engineering, can incorporate a wide variety of arrangements, from working mostly in the office to working mostly from home (e.g., very similar to 100% remote) [39]. In this context, there is a growing concern about implementing strategies to preserve the benefits obtained by software professionals from underrepresented groups and allow them to continue being included in the software development workforce [10].

## 3 METHOD

Case studies are commonly applied in software engineering to investigate a contemporary phenomenon by exploring multiple sources of evidence in a real-life setting, e.g., a software company or software teams [37]. Case studies provide a systematic approach to exploring events, collecting and analyzing real-world data, and reporting findings that can be used to increase the understanding of situations and inform industrial practice [2].

In this research, we conducted a case study because this method is consistent with the phenomenon under investigation, as we are interested in understanding how software professionals from underrepresented groups are experiencing hybrid work. This is a relevant contemporary discussion in the software engineering industrial contexts because there is currently no consensus about how software companies should implement hybrid work, and previous studies have demonstrated that this work arrangement generates conflicting outcomes for professionals and businesses. Regarding this study, it is important to highlight two conceptual aspects:

- *The concept of hybrid work in software engineering*: We are following the definitions of hybrid work proposed by [39], which define this work arrangement as a spectrum that varies from working mostly from home to working mostly from the office, including office mode, remote mode, remote-first, mixed mode, flexible location, among others.
- *Underrepresented groups in software engineering*: Several groups of professionals belong to underrepresented groups in software engineering, including women, Black people, people with disabilities, neurodivergents, and LGBTQIA+ individuals, among others. In this study, we focus on people with disabilities, neurodivergents, and LGBTQIA+ individuals because these are the groups defined as underrepresented in the software company where we collected data.

---

[1] Neurodivergent is defined as having a brain that functions differently from dominant societal norms [19].
[2] Person with disability is defined as an individual who face limations in doing some kind or amount of activities because of ongoing difficulties due to a long-term physical condition [25].
[3] Lesbian, Gay, Bisexual, Transgender, Queer/Questioning, Intersex, Asexual/Allies, and the plus sign is meant to cover anyone else who's not included [21].

Below, we describe this company, i.e., our case study, as well as the strategy we followed to conduct our research. This strategy is based on well-established guidelines defined in software engineering to conduct case studies within the software industry [15, 33, 37].

## 3.1 Selecting the Case

The company selected for our case study was established in 1996 and boasts a robust track record in software development across multiple sectors, including finance, telecommunications, government, manufacturing, services, and utilities. With a workforce exceeding 1,200 professionals, a substantial 70% of them are actively engaged in various software development activities. This group of professionals is organized into over 50 diverse software teams, comprising individuals with a broad spectrum of technical expertise, such as programmers, quality assurance specialists, and designers. The teams also reflect a diversity of individuals backgrounds, encompassing variety of genders and ethnicity.

These software professionals operate within a globally responsive framework, deploying an array of software development methods like Scrum, Kanban, and Waterfall. Their adeptness extends to an understanding of various processes, tools, and techniques, facilitating the creation of cutting-edge software solutions for clients worldwide. The company's clientele spans North America, Latin America, Europe, and Asia, attesting to its global reach and the adaptability of its teams to cater to diverse client needs with efficiency and proficiency.

This company stands out as an ideal setting for investigating post-pandemic hybrid work scenarios due to its current implementation of a dynamic hybrid work structure that aligns seamlessly with the comprehensive definition of the hybrid work spectrum proposed in the literature. Software professionals are afforded the flexibility to navigate across various hybrid configurations, ranging from predominantly remote work to a more office-centric approach, all tailored to accommodate individual preferences and project needs. Additionally, with approximately 1200 software professionals operating in diverse contexts, we can effectively capture the experiences of a diverse cohort of participants, particularly those belonging to underrepresented groups in the software industry.

## 3.2 Data Collection

We applied three data collection techniques: questionnaires, documentary analysis, and observations. Yet, it is worth noting that our *dominant* data collection technique is the questionnaire that gathered both qualitative and quantitative data directly from professionals. The data collected from questionnaires were supported by the data emerging from documents and observations made through official communication channels used within the company, including Slack and email lists. Data collection occurred between April and June 2023. Additionally, it is worth noting that while case studies usually use interviews as the main data collection instrument, we decided to use a questionnaire to collect qualitative data using open-ended questions. This approach aimed to explore the opportunity of understanding the experience of a more diverse cohort of participants, as the company allowed us to interact with all employees in this study.

Regarding this questionnaire, we invited all professionals in the company to answer the questions. However, for this study, we are only interested in the answers from software professionals from underrepresented groups working directly in software development activities, and the answers from other professionals are stored for future work. In particular, while we acknowledge the existence of several underrepresented groups in software engineering, including Women and Black people, in this study, we are specifically focusing on the data collected from three groups of individuals, namely, Neurodivergents, LGBTQIA+, and PWDs, because these are the groups that our case define as minorities.

We used the company's channels to interact with employees and advertise the research, including the official email lists, Slack channels, and WhatsApp groups. While the questionnaire was being promoted, the professionals were constantly reminded that the survey was optional; however, it was important to increase the understanding of hybrid work in their company and the software industry in general. Data collection finished with 722 participants, from which 137 belonged to our targeted population (e.g., software professionals who are neurodivergent, LGBTQIA+, or PWD).

In addition to the questionnaire, for documentary analysis, we collected available records that the Human Resources and IT departments could share about the general amount of people frequently going to the office (e.g., meeting room requests). As for observations, we focused on collecting data from relevant discussions about hybrid work triggered by professionals in the company's channels whenever the subject was mentioned.

*3.2.1 Questionnaire.* To build the survey questionnaire, we first needed to explore how the company is handling post-pandemic hybrid work. Therefore, we interviewed three executives and asked them about their current hybrid work structure and the changes they were trying to implement, e.g., defining the ideal structure based on their business needs.

Based on the answers to this interview, we designed a questionnaire to anonymously collect software professionals' viewpoints on returning to the office on a regular basis. We started the questionnaire with demographic questions about the participants. Then, we asked participants about their work routine and their experience with the voluntarily hybrid work model (i.e., flexible arrangement to decide when/if they are going to the office). We finished the questionnaire by asking participants about their frequency and motives to go to the office lately, their difficulties, and their preferences regarding hybrid work.

A pilot questionnaire was reviewed by a squad created in the company to discuss and understand hybrid work and validated with the participation of five employees. After consideration, some questions were modified or excluded from the questionnaire because our validation process revealed that they could confuse the employees, reduce the number of answers, or review sensitive information about the identity of participants or the project they were involved.

Thus, the questionnaire was modified to limit the closed-ended question about professional roles to include only three answer options (management, design, programming, and QA), as the company does not use other taxonomies for software specializations, e.g., requirements analysts. Further, questions referring specifically to methods and tools used in the projects were excluded since some projects apply tools that, if mentioned, could risk the anonymity of participants and clients. Table 1 presents the full questionnaire used in this research.

**Table 1: Survey Questionnaire**

| Questions | Closed/Open Answer |
|---|---|
| 1. What best describes your gender? | ( ) Female<br>( ) Male<br>( ) Non-binary<br>( ) Prefer not to answer |
| 2. How old are you? | |
| 3. What best describes your ethnicity? | |
| 4. What diversity group do you identify with? | ( ) LGBTQIA+<br>( ) Neurodivergent<br>( ) PWD<br>( ) Prefer not to answer |
| 5. Who shares your household with you? | |
| 6. How many years of experience do you have? | |
| 7. How many years have you been working in this company? | |
| 8. How far is your home from the nearest office, and what is your primary mode of transportation to commute to the office? | |
| 9. What is your primary role in the team? | ( ) Data Science<br>( ) Software Design<br>( ) Software Management<br>( ) Software Programming<br>( ) Software QA<br>( ) Other role |
| 10. Describe an ordinary work week for you. | |
| 11. How often are you going to the office lately? | ( ) I do not go to the office<br>( ) Once or twice a month<br>( ) Once a week<br>( ) 2 or 3 times a week<br>( ) 4 or 5 times a week |
| 12. What are the main reasons that brought you to the office? | |
| 13. Describe how the current work arrangement is affecting you and your life. | |
| 14. Think about a week that you worked mostly from home. What was good and bad that week? | |
| 15. Think about a week that you worked mostly at the office. What was good and bad that week? | |
| 16. Is hybrid work affecting your productivity? Tell us if your overall performance improved, reduced, or remained the same and if you are working overtime. | |
| 17. What is the best hybrid work arrangement? | ( ) 100% remote<br>( ) Voluntarily decide when<br>( ) Weekly pre-scheduled days at the office<br>( ) Monthly pre-scheduled day at the office<br>( ) 100% at the office |

*3.2.2 Documents and Observations.* The data derived from obtained documents and the observation of employee discussions in the company's channels supported us in refining the data collected with the questionnaire and improving our understanding of the actions and behaviors of software professionals from underrepresented groups about hybrid work.

### 3.3 Data Analysis

Both quantitative and qualitative data were obtained in this study. Therefore, due to the nature of these data, we applied two data analysis techniques. First, descriptive statistics [18] were applied to analyze the quantitative data collected from participants in the sample. Second, qualitative analysis was applied to open-ended questions in the questionnaire, document analysis, and notes from the observations based on three coding stages: line-by-line, focused, and theoretical coding [20].

*3.3.1 Quantitative Analysis.* Using descriptive statistics, we focused on quantitatively describing the main characteristics of our sample [18] by partitioning the responses obtained from participants into sub-groups using different statistical functions (e.g., means, proportions, totals, ratios) [15]. Descriptive statistics is widely applied to describe, summarize, and display quantitative data collected in surveys [26].

This analysis method made sense to our closed-ended questions since they were naturally quantifiable. Using descriptive analysis, we organized our sample, the frequency of office visits, and the feelings and preferences of our participants. In this process, we used Tableau[4] to build comparisons among the quantitative data. Tableau is a tool that supports the creation of insightful visualizations from various data sources. We created interactive dashboards with multiple visualizations, including averages and distributions, to help us understand the sample actions and behaviors. We kept changing the visualizations, i.e., exploring different aggregations and combinations within the data, and later, using the qualitative data, we made sense of our case.

*3.3.2 Qualitative Analysis.* Our qualitative analysis process was based on [20] and conducted in three steps:

- *Line-by-line coding*: This step focused on identifying initial codes by analyzing each open-ended answer obtained from the participants to develop concepts within the data. Line-by-line coding allowed us to explore the data using an interactive process, looking closely at what each participant reported to construct codes that reflect individual experiences and perspectives about hybrid work. Figure 1 shows an example of our line-by-line coding strategy.
- *Focused coding*: This step focused on analyzing and reviewing the initial codes, their names, and their content, and by observing existing connections among them, we started creating high-level categories that represent specific aspects investigated in this study, for instance, the problems faced by professionals working in the hybrid environment or what is bringing software engineers to the offices. Figure 2 shows an example of our focused coding process.
- *Theoretical coding*: This step focused on iteratively rearranging the categories until clear definitions of the problem under investigation emerged. At this point, we explored the data to identify how hybrid work affects software professionals from underrepresented groups and how software companies

---
[4]https://www.tableau.com/

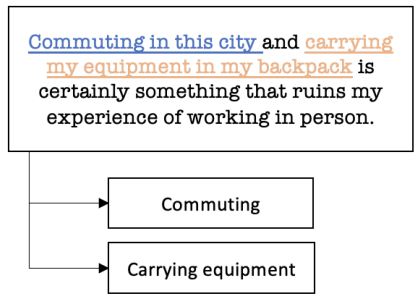

Figure 1: Example of line-by-line coding

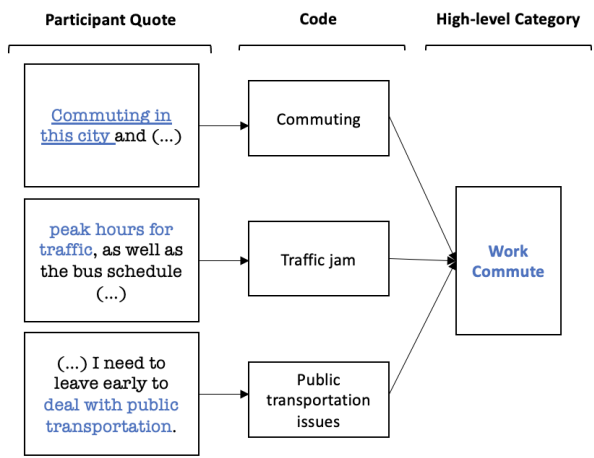

Figure 2: Building categories with focused coding

can support their employees in navigating toward this work model.

The inherent flexibility of this qualitative approach enabled us to dynamically adjust the analysis process in response to the evolving nature of the data, particularly when considering the emergent insights derived from the analysis of quantitative data collected from the case.

### 3.4 Ethics

No personal information about the participants was collected in this study (e.g., name, e-mail, or team) to maintain participants' anonymity. Questionnaires did not require any personal information. We only collected documents that were de-anonymized, and notes and memoing resulting from observing conversations in the company's channel did not include any identification of the employee, only sentences and phrases that were later merged into the pool of qualitative data.

## 4 FINDINGS

In this section, we present our main findings. We start by summarizing the characteristics of our sample of participants. Following this, we present the characteristics of software professionals from each underrepresented group collected in the case. Finally, we discuss the reasons behind software professionals' attitudes towards the voluntarily hybrid work structure.

### 4.1 Demographics

In the overall case study, we collected 722 questionnaires from the whole company and 545 valid questionnaires from software professionals directly engaged in various projects. However, considering the group of participants we were focused on investigating in this study, we received 137 questionnaires from individuals belonging to three underrepresented groups in the case: neurodivergents, LGBTQIA+, and people with disabilities. Looking closely at this group, our participants perform a variety of activities in software development and have experienced hybrid work in different ways. They have an average of 8.1 years of experience working in software development (STD Dev= 6.1 years) and an average of 3.1 years working for the company that participated in the study (STD Dev= 3.3 years). Additionally, almost 42% of the are programmers.

Considering their minority status within the software industry, our group of participants is composed of 46% of individuals who identify as neurodivergent, 47% who identify as LGBTQIA+, and 8% who identify as people with disabilities. Furthermore, beyond the scope of groups defined as underrepresented in the case study, our sample includes 34% of women and 27% of individuals who identify as Black or Mixed-race. Our demographic analysis includes other details such as living arrangements, age, and office attendance. For a comprehensive overview of our sample, refer to Table 2.

Additionally, when examining how our participants embrace voluntary hybrid work, our findings indicate that the majority chose a model closely resembling remote work, i.e., primarily working from home and visiting the office on rare occasions. In contrast, only a small number of individuals established a routine that involves regular office work, such as on a weekly basis, as demonstrated in Figure 3. Therefore, our findings indicate a preference for a hybrid work spectrum closely aligned with fully remote work, regardless of the underrepresented group to which software professionals belong. In the next section, we explore the distinct reasons and motivations guiding professionals from each group in their choice of hybrid work environments.

### 4.2 Hybrid Work: Six Reasons Why Software Professionals from Underrepresented Groups Visit the Office

Looking at our whole sample, software professionals from underrepresented groups attend the office for six prominent reasons: social activities, management requests, team meetings, technical support, infrastructure needs, and client demands. At this stage, it is noteworthy that despite professionals operating within a voluntary hybrid work model, allowing them the flexibility to choose their office attendance, our participants identified reasons that are more project-driven than solely optional in nature.

Approximately 28% of our participants indicated that *social activities* serve as their primary motivation to work in person. These activities may include informal gatherings with colleagues (e.g.,

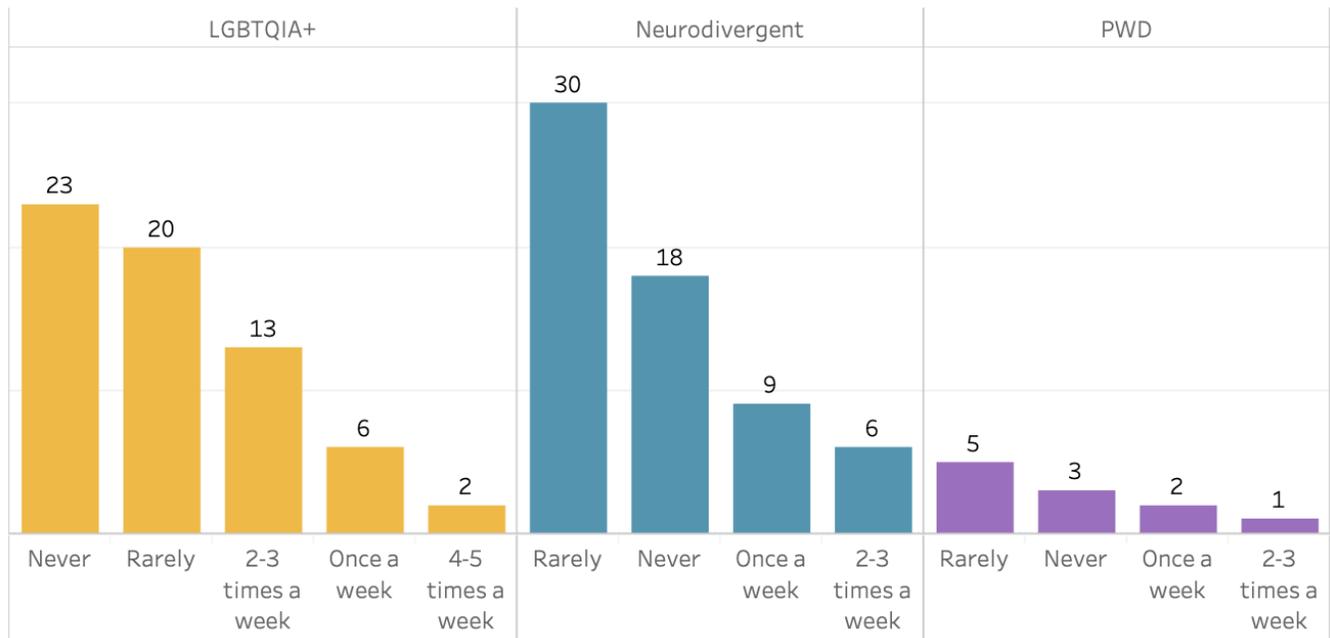

Figure 3: Office Attendance

Table 2: Demographics

| | Participants Profile | |
|---|---|---|
| Gender | Male | 63% (86 individuals[a]) |
| | Female | 34% (47 individuals[b]) |
| | Non-binary | 2% (3 individuals) |
| | Prefer not to answer | 1% (1 individual) |
| Age | 18-25 years old | 22% (30 individuals) |
| | 26-35 years old | 60% (82 individuals) |
| | 36-50 years old | 17% (23 individuals) |
| | 50+ years old | 1% (2 individuals) |
| Ethnicity | White | 70% (97 individuals) |
| | Mixed-race | 17% (24 individuals) |
| | Black | 10% (13 individuals) |
| | Asian | 1% (1 individual) |
| | Prefer not to answer | 1% (2 individuals) |
| Diversity Group | Neurodivergence | 46% (63 individuals) |
| | LGBTQIA+ | 47% (64 individuals) |
| | PWD | 9% (11 individuals) |
| Living Arrangement | Living with Family | 77% (105 individuals) |
| | Living Alone | 20% (27 individuals) |
| | Living with Friends | 4% (5 individuals) |
| Role in Software Project | Software Programming | 42% (57 individuals) |
| | Software QA | 25% (34 individuals) |
| | Software Design | 23% (32 individuals) |
| | Software Manager | 6% (9 individuals) |
| | Data Scientist | 4% (5 individuals) |
| Years of Experience | 0-5 years | 40% (55 individuals) |
| | 6-10 years | 32% (44 individuals) |
| | 11-15 years | 18% (24 individuals) |
| | 16+ years | 10% (14 individuals) |
| Time in the Company | 0-1 years | 44% (60 individuals) |
| | 2-5 years | 44% (60 individuals) |
| | 6-10 years | 8% (11 individuals) |
| | 11+ years | 4% (6 individuals) |

Notes: [a] 2 transgender men. [b] 2 transgender women.

coffee or lunch), company-sponsored workshops and learning activities, or annual events such as holiday celebrations.

*Management requests* are the second primary factor driving professionals to the office, and it is encountered by 20% of our sample. These requests are not entirely voluntary; instead, they stem from a management directive by software project managers or team leaders. This requirement is based on the necessity of establishing a shared routine for all team members, such as incorporating core hours when everyone must be present at the office.

Moreover, 13% of our sample visit the office whenever they encounter technical problems with their equipment that cannot be addressed remotely, triggering the need for *on-site technical support*. This support includes both hardware and software issues, along with administrative paperwork that requires in-person handling, making this a less voluntary and more mandatory reason.

Additionally, 12% of participants in our case study said that *infrastructure needs* prompts their presence in the office. These infrastructure needs manifest as a multifaceted motivation, as they may involve employees voluntarily seeking on-site perks like a better internet connection, air conditioning, or the convenience of the office location (e.g., proximity to a service needed by the employee). However, infrastructure can also refer to the existence of a structure (hardware or software) that is accessible only on-site, compelling professionals to be in the office to fulfill their tasks.

Following this, 11% of the participants attend the office for *team meetings*. Unlike managerial requests, these meetings are more voluntary in nature, driven by a collective sentiment that the team should gather to discuss project aspects or address issues that are challenging to resolve when professionals are dispersed across different locations. Hence, each individual determines whether their presence is required for such a meeting or not.

Finally, around 4% of participants work on-site due to *client needs*. In this particular situation, contractual agreements mandate that the work must be completed on-site. Consequently, to meet the demands of clients, professionals are expected to be in the office more frequently than they might prefer.

It is important to note that while we have identified six primary reasons motivating software professionals from the underrepresented groups examined in this research work on-site, the dynamics of these factors change significantly when each group is analyzed independently. Our findings indicate that social activities emerge as the primary motivator for neurodivergent software professionals, while LGBTQIA+ individuals tend to frequent the office in response to management requests, and for individuals with disabilities, attendance is triggered by the need for technical support. A detailed breakdown of these findings is presented in Table 3.

Table 3: Reasons Behind Professionals Attendance

| Reason | Neurodivergent | LGBTQIA+ | PWD |
| --- | --- | --- | --- |
| Social Activities | 28% | 0% | 0% |
| Management Request | 4% | 16% | 0% |
| Technical Support | 5% | 0% | 8% |
| Infrastructure | 0% | 12% | 0% |
| Team Meeting | 11% | 0% | 0% |
| Client needs | 0% | 3% | 0% |

## 4.3 Hybrid Work Affects Software Professionals Focus, Engagement and Health

Our findings demonstrated that *focus* is the primary concern of software professionals from underrepresented groups in hybrid work environments. According to the narratives of our participants, they have learned to navigate and manage concentration in their home offices over time (e.g., sharing the space with family members). Consequently, they now experience more interruptions and distractions when working in the office. Particularly, prompted by the loud and constant conversations within the office environment, these professionals often seek a quiet space at home to concentrate on the software project activities. It is noteworthy that neurodivergent professionals, in particular, express a greater interest in how working from home supports them in maintaining focus and avoiding distractions.

*Work commute* stands as another crucial factor in the decision-making process for software professionals when choosing between working from home or at the office in hybrid environments. In this context, our findings suggest that for software professionals from underrepresented groups, the concern goes beyond the time spent in transit between the two environments and encompasses health and safety issues. Frequent commuting can potentially increase the levels of anxiety, stress, and exhaustion, especially for neurodivergent and LGBTQIA+ individuals. In the case of LGBTQIA+ individuals, the fear of discrimination in public spaces often leads them to conceal their identities to be in the office, which leads them to divide their focus between how they present themselves to others and the tasks they need to perform.

Hybrid work plays a crucial role in supporting professionals' *health and work-life balance*, especially those from underrepresented groups. This entails providing flexible work hours for managing personal activities, encouraging support from family members, and allowing the adjustment of work schedules to individual needs. This adaptability is particularly beneficial for people with disabilities and neurodivergents, as it facilitates the implementation of personalized accommodations tailored to their specific requirements. Additionally, our participants highlighted that the flexibility provided by hybrid environments supports their medical treatments and leads them to healthier routines.

Participating in *social interactions* is a key element of a hybrid work routine, enabling software professionals to adjust the degree of engagement with co-workers according to their preferences. For some professionals, prolonged periods of remote work may lead to feelings of boredom and isolation, while regular visits to the office foster a sense of belonging, encourage collaboration among peers, and contribute to professional development. Yet, for some software professionals from underrepresented groups, increased social interactions not only disrupt focus and concentration but also diminish their privacy and comfort.

In a broader context, our findings suggest that individuals from some minorities in the software industry, specifically neurodivergent, people with disability, and LGBTQIA+ individuals, are more likely to find satisfaction in the prospect of working in hybrid environments that allow them to predominantly work from home. This setup has provided them with the levels of concentration, safety, privacy, and health support required to meet their needs. Additionally, they can manage their social interactions with peers, enabling them to navigate the collaboration necessary to meet project requirements and adjust their sense of belonging accordingly. Table 4 provides evidence for each factor presented in this section, which is derived from quotations provided by participants.

## 4.4 Hybrid Work and Perceived Performance

Software professionals attribute increased productivity to the flexibility of choosing their work location. As reported in the previous section, individuals from minority groups, such as neurodivergent software professionals, face numerous challenges concentrating in the office. Additionally, the stress and anxiety caused by several factors associated with commuting between home and the office contribute to a perception of increased productivity when working remotely.

Concerning perceived productivity, we directly inquired with our participants about their self-assessment of performance within a hybrid work environment, and more than 60% of them perceived a productivity increase when working from home. In contrast, 9% of the sample observed a decrease in their productivity. Yet, almost half of the sample indicated working overtime to complete their tasks.

More specifically, as illustrated in Figure 4, although a significant portion of our group reports an increase in productivity, they also engage in a substantial amount of overtime during the week. Our analysis indicates that the reasons behind this overtime are inconclusive, prompting various hypotheses, such as doing overtime to avoid office visits due to low performance, blurred boundaries between work and home, or problems that professionals are ignoring and not reporting. Further evidence is essential to conclusively understand how hybrid work influences the perceived performance of software professionals, especially those from minority groups. Yet, our findings indicate that professionals are happy with a hybrid work structure that predominantly embraces remote work.

Table 4: Example Evidence from Participants Quotes

| Category | Definition | Example Evidence |
| --- | --- | --- |
| Focus | Maintaining focus proved to be challenging in the office nowadays, prompting professionals to prefer concentrating on their tasks while working on their personalized and familiar environments at home. | -"my visits to the office were quite specific, and undoubtedly, the noise and talking around diminished my focus rather than assisted me [with my tasks]" (P43)<br>-"since I have ADHD, working at home allows me to have fewer distractions compared to the office, which significantly increases my productivity." (P71)<br>-"I have difficulties focusing in environments where people are talking among themselves, coming and going from the premises, and not even my prescribed medications are helping me with this." (P75)<br>-"as a neurodivergent person, I felt that I could concentrate better and feel more comfortable, as well as be more productive without some of the attention-disrupting factors in the office." (P128)<br>-"In general, I feel mentally better not having to interact obligatorily with everyone, engage in small talk, or have a call or conversation happening beside me when I'm trying to concentrate." (P069) |
| Work Commute | The time required to commute between home and the office demotivates software professionals, which in many cases increases anxiety, stress, and fatigue, leading to high levels of exhaustion, ultimately. | -"due to the time and effort dedicated to commuting, as well as being away from family, in situations where I needed to work in person, I experienced increased anxiety and fatigue" (P82)<br>-"another negative factor is the constant fear of danger; I have been robbed before, and just walking on the streets [with the equipment] caused me anxiety." (P58)<br>-"likely more tiring, as it requires me to wake up earlier and get all dressed to go." (P50)<br>-"In a week of work, I lost 17:30 hours, almost a whole day just in commuting to and from work." (P84)<br>-"I don't need to worry about going out for lunch, traffic on the way to or from work, the risk of being robbed and having my work computer taken, it eases my anxiety being at home." (P95) |
| Health and Work-Life Balance | Software professionals discover enhanced approaches to balance their work and personal activities, including handling medical appointments, thanks to the increased flexibility of working from home more frequently than going to the office. | -"the increased flexibility in my work hours has been helping me with important matters such as adhering properly to the medical treatments I need." (P82)<br>-"Working from home allows me to bypass the stress of daily commuting and use the extra time to engage in physical activities, study, and spend time with my family." (P13)<br>-"[I have] more time to spend with my family, and it is easier to attend medical appointments." (P91)<br>-"This model gives me the flexibility to exercise during lunchtime or before starting my workday, which helps my mental health a lot!" (P91)<br>-"I struggle with depression and often experience shifts in energy levels, so working from home allows me to better navigate these periods of low energy, as I can organize my activities based on how I'm feeling." (P07) |
| Social Interactions | Hybrid work accommodates diverse social preferences among software professionals, allowing personalized engagement. For neurodivergent, LGBTQIA+, and PWD individuals, it fosters interaction with peers while respecting individual boundaries, and enhancing overall well-being. | -"The greatest benefit of working in the office is being closer to the team, seeing people in person." (P110)<br>-"[in the office] people feel more comfortable to speak and ask questions. For example, I don't feel I can play music and focus solely on my work because people interact with me at any time." (P137)<br>-"increased contact with the work team for immersion and achievement of one specific goal for that day." (P104)<br>-"it can be uncomfortable [being around people] when you are having menstrual cramps." (P97)<br>-"nowadays, I go to the office when I want to enjoy the company of people I already know or when I want to create connections and develop new interactions." (P108) |

### 4.5 Software Professionals from Minority Groups Prefer Hybrid

We inquired professionals about their preferences for an ideal hybrid work structure that works for individuals from underrepresented groups, and 66% of our sample favors a flexible hybrid environment, allowing them to work predominantly from home and decide voluntarily when to be in the office. Additionally, 20% would prefer a fully remote work structure. Another segment, accounting for 12% of the sample, prefers a hybrid structure with mandatory weekly office attendance. Meanwhile, 3% find a mandatory monthly office visit sufficient for their needs.

None of the individuals in the sample indicated a preference for returning to daily or predominant on-site activities, such as meeting at the office every day. This outcome suggests that, despite the challenges of working from home for some software professionals from minority groups, overall, these professionals have adapted to these challenges, and now, they anticipate the continuation of hybrid work environments in the software industry, providing support for their individual limitations and enabling them to keep contributing to a diverse workforce in the field.

## 5 DISCUSSION

The initial goal of this research was to investigate how hybrid work affected software professionals from minority groups, in particular, neurodivergent individuals, LGBTQIA+ people, and PWD. Our investigation leads to an important outcome: voluntary hybrid work is important for professionals from underrepresented groups.

### 5.1 Hybrid Work Helps Neurodivergent Software Professionals to Cope with Work and Health Challenges

Neurodivergent individuals often confront workplace challenges arising from a lack of understanding about their conditions, which may lead to discrimination. Moreover, inadequate work settings that do not accommodate their needs can result in performance issues and health concerns [9, 30]. In the software industry, significant stress levels are observed among neurodivergent groups of practitioners [23].

Hybrid work is demonstrated to be well-suited to support neurodivergent software professionals, offering a setup that respects personal boundaries and allows for necessary individual adjustments. Specifically, the hybrid environment supports these professionals in focusing on highly concentration-demanding software activities, helping them avoid noisy surroundings, frequent interruptions, and distractions while managing stress and anxiety levels. Ultimately, individuals undergoing medical care find the hybrid setting ideal for coping with their treatment appointments without significant interference in their work routine or the dynamics of their software team.

### 5.2 Hybrid Work is Safer and More Inclusive for LGBTQIA+ Software Professionals

LGBTQIA+ individuals often face unique concerns in the workplace, including worries about acceptance and apprehension regarding the potential impact of their identity on career progression and relationships with co-workers [11, 14]. In traditional work settings, these issues are often exacerbated, especially in a heterosexual male-dominant field like software engineering, prompting LGBTQIA+ software professionals to fear discrimination and harassment, hence investing extra effort in conforming to perceived norms in appearance and behavior [35, 36].

Hybrid work emerges as an ideal work setting for LGBTQIA+ software professionals, allowing them to navigate the engagement and social interactions required to work on software projects at their own pace. They can initiate interactions remotely and seamlessly transition to the office setting when they feel most comfortable. Additionally, hybrid structures afford these professionals the individual privacy they need, particularly in software companies not fully adapted to the diversity of the LGBTQIA+ population (e.g., transgender software professionals).

### 5.3 Hybrid Work provides PWDs in the Software Industry an Adaptable and Personalized Work Environment

In general, individuals with disabilities often face diverse physical and mental challenges at work, including insufficient accessibility, inadequate settings, hostile environment influenced by social stigma, rigid work schedules, and health-related issues originating from workplace pressure and inadequate organizational support [42, 46]. Some of these challenges are also experienced by PWDs working in software development activities [31].

Hybrid work environments empower software professionals with disabilities to tailor their workspace to their specific needs. In contrast, on-site workstations, while adapted for accessibility, often target the diverse needs of a group simultaneously, which can potentially limit the individual work experience. Therefore, the flexibility of hybrid structures facilitates diverse setup options, enabling people with disabilities to plan and prepare according to their unique requirements and the demands of the software project.

### 5.4 Actions for Inclusiveness

Our study highlights the advantageous impact of hybrid work environments on software professionals from minority groups—especially PWDs, neurodivergents, and LGBTQIA+ individuals. Our analysis reveals the positive effects of this practice on adaptability, flexibility, concentration, engagement, and work-life balance. However, we also identified red flags that call for attention from software organizations to foster inclusiveness in the field.

Hybrid work enhances focus and concentration by allowing software professionals from minority groups to work from home when needed. However, the absence of incentives to visit the office and interact with others can lead to isolation. In this sense, we highlight the need for software organizations to implement proactive strategies that encourage in-person socialization and foster virtual collaboration, team-building, and a sense of community. Neglecting these aspects may negatively impact inclusiveness in software engineering.

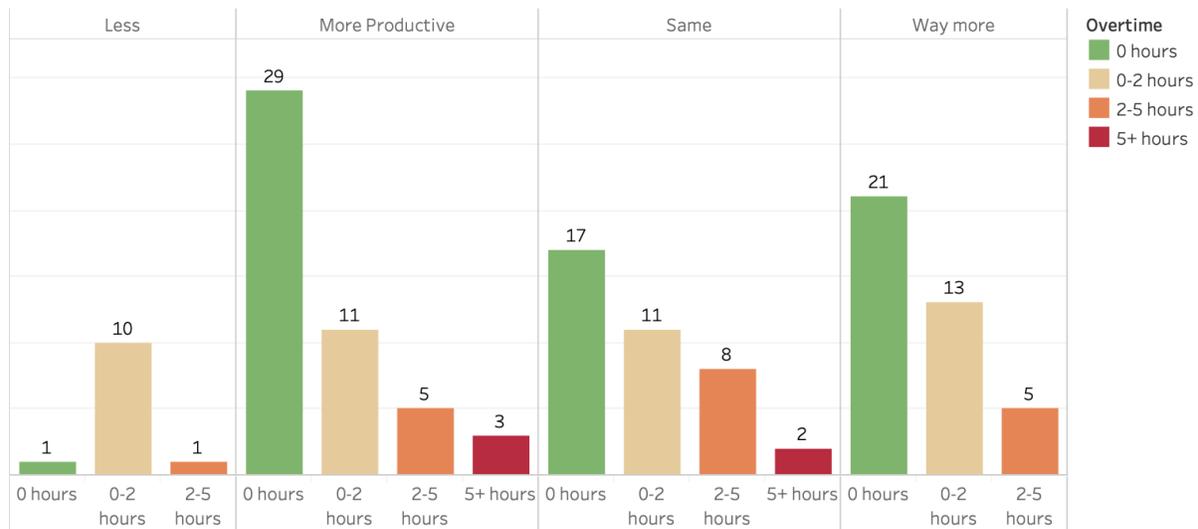

Figure 4: Voluntarily Overtime

Additionally, we observed that infrastructure is a key reason driving professionals to the office, especially among software engineers with disabilities. While this may not seem problematic on a broader scale, within the specific group we are analyzing, it raises concerns, particularly because PWDs were the leading group reporting this problem. In this sense, we strongly recommend that software companies enhance their infrastructure services for professionals in hybrid structures to prevent exclusion and disparities, which could significantly impact individual productivity.

From a broad perspective, hybrid work is generally advantageous for software professionals from underrepresented groups. While certain limitations exist and could potentially cause substantial harm, software organizations have the ability to intervene and implement strategies to mitigate these issues. Overall, hybrid work stands out as a promising practice to enhance diversity and inclusion in software development environments, affording professionals from minority groups the flexibility and adaptation necessary to address previous challenges encountered in traditional office settings.

## 5.5 Limitations and Generality of Results

Our research has methodological limitations inherent in case studies. Firstly, our findings are subject to our interpretation of the data collected from the field. Our strategy to mitigate this threat to validity was using a triangulation approach by collecting data from various sources. Additionally, we strongly relied on the quotations provided by the participants to constantly compare and contrast codes and themes emerging in our analysis.

Secondly, as is common in case studies, our findings are not generalizable in a statistical way (e.g., to a broader population). Instead, we claim analytical generalization, which is the most common way of generalizing findings in qualitative studies and involves extrapolating findings from a specific case or context to a broader theoretical understanding. In this sense, we expect that researchers and practitioners can draw insights from our discussions and transfer the knowledge acquired from our case study to their unique situations and contexts. To allow for this analytical generalization, we provided contextual details about the case, the context, the participants involved in the study, and anonymized quotations extracted directly from the participants.

Ultimately, it is worth noting that our study is limited in providing tailored recommendations to individuals within the minority groups we investigated. For instance, ethical constraints prevented us from directly inquiring about participants' specific neurodiversity (e.g., autism, ADHD, dyslexia, etc.), sexual orientations (e.g., lesbian, gay, bisexual, etc.), or disabilities. Consequently, our discussions and recommendations are based on a collective perspective. However, we acknowledge the individuality of each participant in our case and aim for more in-depth investigations to address these unique characteristics within each group in future research.

## 6 CONCLUSION

After COVID-19, many software companies embraced hybrid work structures. This study explores the experiences of underrepresented groups in software engineering adjusting to this model, specifically neurodivergents, LGBTQIA+ individuals, and PWDs. Our study occurred at a large South American software company that implemented a voluntary hybrid work model. Our general findings suggest a common preference for a hybrid work spectrum closely aligned with predominantly remote, regardless of the minority group to which the professional belongs.

In the post-pandemic era, these individuals prefer the tranquility of home offices over traditional office settings, citing noise and distractions as detriments to their focus. The flexibility of the hybrid model allows for personalized work setups, improving well-being and accommodating health treatments more efficiently. However, the lack of incentives for social interactions raises concerns about the potential isolation of these professionals, demanding proactive strategies for both in-person and virtual socialization. Additionally, infrastructure support, particularly significant for professionals with disabilities, requires improvement to prevent exclusion and disparities that may lead to low productivity.

Overall, hybrid work stands as a promising strategy for fostering diversity and inclusion in the software industry, addressing past limitations faced by underrepresented groups in traditional office settings. Our study lays the groundwork for future research, including proposing strategies to enhance socialization and foster a sense of belonging in hybrid environments, as well as in-depth investigations that consider the particularities inherent in different underrepresented groups in software engineering. We expect that our study represents a significant step toward advancing a more diverse and inclusive landscape in the field.